\documentstyle[subeqn,preprint,aps]{revtex}
\def\half{\textstyle{1\over2}}

\begin{document}
\preprint{EHU-FT/9704, UG-1/97, DTP/97/3, hep-th/9705001}
\draft
\renewcommand{\topfraction}{0.8}
%%%%%%%%%%%%%%%%% HERE comment out next 2 lines
%@twocolumnfalse\endcsname
 
\title{SELECTION RULES FOR SPLITTING STRINGS}
\author{Ana Ach\'ucarro$^1$ and Ruth Gregory$^2$}
\address{\  \\
${}^1$ Departamento de F\'\i sica Te\'orica, 
Universidad del Pa\'\i s Vasco, Lejona, Vizcaya, Spain.\\
Department of Theoretical Physics, University of Groningen, The Netherlands.\\
Department of Mathematics, Tufts University, Medford, Massachusetts 02155.\\
\ \\
${}^2$ Centre for Particle Theory, 
Durham University, South Road, Durham, DH1 3LE, U.K.
}
\date{\today}
\maketitle
\begin{abstract}
It has been pointed out that Nielsen-Olesen vortices may be able to
decay by pair production of black holes.  We show that when the
abelian Higgs model is embedded in a larger theory, the additional
fields may lead to selection rules for this  process - even
in the absence of fermions - due to the failure of a
charge quantization condition. 
We show that, when there is topology change, the criterion 
based on the charge quantization condition supplements the 
usual criterion based on $\pi_0(H)$. In particular, we find that, 
unless $2\sin^2\theta_W$ is a rational number, the  thermal splitting 
of electroweak Z- strings by magnetically neutral black holes 
is {\it impossible}, even though $\pi_0(H)$ is trivial.

\end{abstract}
\pacs{PACS numbers: 04.70.-s, 04.70.Dy, 11.27.+d \hspace*{2cm} 
hep-th/9705001}

%%%%%%%%%%%%%%%%% HERE comment out next  line
%\vskip2pc]

%%%%%%%%%%%%%%%%%%%%%%%%%%%%%%%%%%%%%%%%%%%%%%%%%%%%%%%%%%%%%%%%%%%%%% 
Topological defects and  solitons remain an active field of study
in many different areas of physics. Generally, as is implied
by the nomenclature, defects are argued to be stable from
topological considerations of the vacuum structure of the particular
theory in question. Recently, however, it was observed that supposedly
stable vortices in the abelian-Higgs model may be able to decay
by pair production of black holes\cite{EHKT,HR,E1}. 
The instanton describing this
process is based on the  C-metric \cite{KW}, or the Israel-Khan (IK)
metric \cite{IK} for thermal nucleation of uncharged black holes 
\cite{GH,E2}, where
the conical singularity in the exact general relativistic solution
is smoothed out by a real vortex with the same deficit angle.
Briefly, the C-, and IK-metrics represent two black
holes connected to infinity by conical singularities. For the C-metric
these black holes are undergoing uniform acceleration away from each other 
and are not in causal contact. The IK-metric on the other hand is
static, with the two black holes held in equilibrium by the conical
deficits, 
whose tension exactly balances
the gravitational attraction between the black holes.
The spatial topology of these solutions is 
$S^2\times S^1$ minus one (C) or two (IK) points.
The euclideanization of these
metrics gives the instanton which interpolates between the initial,
infinite vortex, state and the final, split vortex, state. This 
process clearly involves a topology change in the spacetime, thus
circumventing the usual stability arguments for the topological defect,
which rely on spacetime being topologically trivial.

Nonetheless, topology change is not the only ingredient in this decay
process; the decay of idealized strings (conical deficit sources) via
purely gravitational processes has been commented upon in the
literature some years ago \cite{GP}. Central to this particular
construction of a field-theoretic vortex splitting is the
demonstration \cite{GH,AGK} that an abelian-Higgs vortex \cite{NO} can
emanate from a black hole, and the solution falls outside the scope of
several well known ``no hair" theorems for abelian Higgs fields
\cite{APLMB}.  While in flat spacetime there is no regular field
configuration describing a topological string with ends, this is not
true when there are black holes present.  The existence of
non-contractable spheres in spacetime (due to the black hole horizon)
allows for the vortex to end in a black hole \cite{EHKT,AGK} in such a
way that the field configuration is regular outside the horizon.  In
effect, the field configuration is that of a magnetic monopole - the
only difference being that the vortex lives in a phase with broken
symmetry where electromagnetism is massive, causing the flux lines to
be confined to the vortex core.  It is well known that Dirac monopoles
\cite{D} obey a charge quantization condition
\begin{equation}
\label{dirqn}
eg = \hbar/2
\end{equation}
which is rather elegantly described in terms of the Wu-Yang construction
\cite{WY}. The cosmic string also obeys a flux 
quantization condition coming from
single-valuedness of the Higgs field which forms it, and the coincidence
of these quantization conditions allows us to use a Wu-Yang construction
to achieve a smooth field configuration for the instanton.

We can often embed the abelian Higgs model in a larger theory to
obtain classically stable (topological or non-topological) strings
whose energy momentum tensors are identical to those of the
Nielsen-Olesen vortex,
hence they can also
avoid no-hair theorems and may in principle split by pair creation of
black holes.  This was somehow implicit in \cite{EHKT,HR,E1}, where
regularity of the instanton required a second U(1) gauge field (with
respect to which the Higgs field was neutral).  A less trivial example
is the semilocal string \cite{VA} which occurs in a special limit of
the Weinberg-Salam model where the SU(2) symmetry is global.
Even
though the string is not topological (because the vacuum manifold,
$S^3$, is simply connected) it has been shown that whenever $m_{_{\rm
Higgs}} \leq m_{_{\rm gauge}}$, infinitely long, straight strings are
stable both to small perturbations\cite{VA,AKPV,P} and to semiclassical
tunnelling in flat space \cite{P,PV}. When the $SU(2)$ coupling is
non-zero, electroweak Z-strings are also classically stable in a very
small region of parameter space very close to that of stable semilocal
strings \cite{V,JPV}, but they can decay by nucleation of magnetic
monopoles \cite{P,PV,N}.  The classical and 
semiclassical stability of more general
embedded vortices in flat spacetime has been investigated in
\cite{BBV} and \cite{P,PV}, respectively.

Here we are interested in semiclassical decay by {\it spacetime}
topology change -- which can obviously affect both topological and
non-topological defects, so we will concentrate on the former. The
tunneling amplitude found in \cite{EHKT,HR,E1} for Nielsen-Olesen
vortices was incredibly small, of order $e^{-10^{12}}$ for GUT
strings, however, it does provide the only possible decay channel for
otherwise stable defects and it is interesting that this channel can
{\it close} when the field theory model is extended.  It is worth stressing
that the arguments we will use make no reference to fermions. It is
well known that the presence of fermions can
jeopardize instanton-mediated processes due to incompatibility of
spin structures \cite{WGH}.
However in a purely bosonic extension of the theory one
would expect the decay mode to persist, since the instanton will still
be a solution to the (Euclidean) field equations with all extra fields
set to zero.  In fact, in \cite{KLW} it was shown that the presence
of additional symmetries should typically enhance the decay rate of a false
vacuum by bubble nucleation due to the extra zero modes.  By contrast,
the decay of vortices by black hole pair creation seems to suffer from the
opposite effect: there may be selection rules for the process
precisely due to the extra bosons.

Let us now consider the conditions for
the splitting of a string. Recall that the abelian-Higgs vortex could
split because it was possible to use a Wu-Yang construction to
smoothly convert from the vacuum configuration exterior to the string
($\Phi = \eta e^{i\phi};\ A_\mu = {\textstyle{1\over e}} \nabla_\mu
\phi$) to the `true' vacuum ($\Phi = \eta;\ A_\mu =0$) using the gauge
transition function ${\cal G}(\phi) = e^{i\phi} \in U(1)$ on an
equatorial region in which two distinct gauge patches overlap. Now
consider what happens if we extend our original model by the addition
of extra bosonic fields.  If these extra fields couple to the original
U(1) gauge field, or indeed if there are additional gauge fields
(e.g.\ if the $U(1)$ becomes a subgroup of a larger symmetry group),
then they too will have to be defined separately on each gauge patch,
and transformed by the gauge transition function according to the
particular representation in which they lie. A necessary condition for
the string to split would therefore seem to be that these field
configurations can all be consistently defined.  One can view this as
a relic of the Dirac charge quantization conditions (DQC's) in the
unbroken theory. A string ending on a black hole represents confined
magnetic flux from a magnetically charged black hole created prior to
a phase transition.  Such a charged black hole must be made up of
particles in the spectrum of the theory, which are known to obey
quantization conditions analogous to eq.\ (\ref{dirqn}).  Therefore
{\it a string can terminate on a black hole (or monopole) only if
their combined (confined plus unconfined) flux corresponds to some
magnetic monopole in the symmetric phase}. (Note that 
this condition must
hold over and above any other conditions necessary for the
regularity of the instanton \cite{EHKT,HR}, which we assume are
satisfied in what follows.)

Clearly this criterion also applies to the case of a non-topological
vortex ending in a regular monopole, where it is customary
to state the condition in terms of properties of the unbroken
residual symmetry group, $H$.  We shall do this now for the black hole
case, making reference to two specific examples, discrete gauge
symmetry and the electroweak string, in order to illustrate: a) that a
commonly used relation between $\Pi_0(H)$ and Aharonov-Bohm phases is
not strictly correct when there is gauge mixing, and b) that neither Aharonov-Bohm
interactions nor $\pi_0(H)$ are enough to determine when a string can
split and one should always look at the DQC's as well.

Given a symmetry breaking $G \to H$ which admits localized vortices 
with finite energy per unit length, parallel transport of the Higgs 
field around the
string implies
\begin{equation}
\label{pint}
{\bf P} {\rm exp}\{ i \oint {\bf A}_{\rm H} \} \Phi  \equiv {\bf g} 
\Phi = \Phi 
\end{equation}
where ${\bf A}_{\rm H} = e_{\rm Higgs} A^a T^a_{\rm Higgs}$;
i.e.\ the group element {\bf g} 
must sit in the unbroken group $ H$. 
To derive a condition for the splitting of a string, we must consider
what happens to {\bf g} as we deform a loop encircling the string,
slipping it off the end of the string and contracting it to a point.

As long as there is no patching involved, it is clear that
infinitesimal transformations of the loop affect {\bf g} continuously,
and that when the loop is contracted to a point the final value of
{\bf g} must be the identity. In that case a necessary condition for
the string to split is that {\bf g} be continuously connected to the
identity - the usual criterion for a (non-topological) string to end
in a regular monopole.  The only subtlety,  when a string
ends in a black hole, is how patching conditions may affect {\bf g}.
If $H$ is abelian, {\bf g} is unchanged; if $H$ is not abelian, {\bf
g} is not uniquely defined but depends on a reference point on the
closed loop (all physically measurable quantities are, of course,
independent of such a reference point).  Under a gauge patching, {\bf
g} will be conjugated by the transition element at that point, which
is well defined because the transition function must be single valued
in $G$ (this is the DQC). Once in the ``vacuum'' patch we are back to
the previous case: the endpoint value is the identity, and the
conjugated {\bf g} must be continuously connected to it; but, since
the identity is fixed under conjugation, and conjugation is a
continuous function, this statement is also true in the other
patch. We conclude that a {\it necessary} condition for a string to
split, whether it is by black holes or monopoles, is that {\bf g} lie
in the trivial component of $\pi_0(H)$.  This condition was given in
refs. \cite{EHKT,HR} with an acknowledgment to a private communication
by J. Preskill. However the condition is not {\it sufficient}, as we
will now show.

In physical terms, an element {\bf g} not equal to the identity simply
indicates a nonzero magnetic flux emanating from the monopole or black
hole.  If one demands that the string split by nucleating {\it neutral}
black holes 
(as required by the Israel-Khan instanton or in a theory with {\it no} residual electromagnetic symmetry) 
then the condition {\bf g}=1 is not only {\it necessary} but {\it 
sufficient}.  Indeed, the vortex
corresponds to an element of $\pi_1(G/H)$; if {\bf g}=1, the loop in
$G/H$ is closed in $G$ and one can simply use the `inverse loop' as a
transition function to make the string disappear into a Schwarzschild
black hole. Therefore, if ${\bf g} \neq 1$, thermal splitting of a 
string by nucleation of magnetically neutral black holes will not 
take place even if ${\bf g}$ is continuously connected with the identity. 

Let us illustrate this rule with two specific examples.
The first one, also mentioned in \cite{EHKT}, is the coupling of gravity to an abelian Higgs
model with an extra, fractionally charged field. This model has been
extensively studied in connection with (discrete) quantum hair for
black holes \cite{CPW}.  
The matter part of the action is:
\begin{equation}
\label{dqh}
S = \int d^4x \sqrt{ g} \biggl[
 |D_\mu \Phi|^2 + |D_\mu \chi|^2 -{\textstyle{1\over 4}} F_{\mu\nu}^2 
- V(\Phi, \chi) \biggr]
\end{equation}
where $D_\mu \Phi = (\nabla _\mu -iNeA_\mu)\Phi$, \ \  
$D_\mu \chi = (\nabla_\mu -ie A_\mu)\chi$ and $V(\Phi, \chi)$ contains
a mexican hat potential for $\Phi$. Under a U(1) gauge transformation,  
\begin{equation}
\label{gt}
\Phi \to   e^{ i Ne \alpha }\Phi  , \qquad     
\chi  \to   e^{ i e \alpha }\chi  , \qquad   
A_\mu \to A_\mu + \partial_\mu \alpha  , 
\end{equation}
thus $\alpha$
is to be identified with a period of $ {2\pi / e}$. If the Higgs
field $\Phi$ condenses, this results in a symmetry breaking $U(1) \to
{\bf Z}_N$, topological vortices can form and although they are classified by 
$\Pi_1(G/H)={\bf Z}$, they also can be labelled by $\pi_0(H) = 
{\bf Z}_N$, which characterises
their Aharonov-Bohm interactions with the fractionally charged $\chi$
quanta. Applying the $\pi_0(H)$ criterion, it is 
obvious that strings whose winding number is
not a multiple of N cannot break by nucleating black holes, because
{\bf g}$\neq$1 for such strings and $H$ is discrete.
It is interesting that for the string itself, the putative gauge
transition function ${\cal G}(\phi) = e^{i\phi}$ or $\alpha(\phi) = 
\phi/Ne$, would `unwind' the vortex, but it is not legal
since it is not closed in $G$ (as can be seen by its non-single
valuedness acting on $\chi$). This is clearly equivalent to the DQC
in the symmetric phase, since the presence of particles of fundamental
electric charge $N$ times smaller than that of $\Phi$, gives a 
minimal magnetic monopole charge $N$ times larger. Thus, both criteria 
are equivalent in this case.

Our second example is a `pruned' electroweak string. Consider a symmetry
breaking 
U(1)$\times$U(1)/${\bf Z}_2\to$U(1), realised by two 
Higgs fields $\phi_1,
\phi_2$ which have equal charge $g/2$ with respect to the first U(1), but
opposite charge $\pm g'/2$ with respect to the second U(1). Writing
$\Phi = ({\phi_1\atop \phi_2})$ for brevity, the
gauge covariant derivative is 
\begin{equation}
D_\mu \Phi = \left [ \nabla_\mu - {\half} i g' B'_\mu \sigma^3 - {\half} i g
B_\mu \right ] \Phi \ \ .
\end{equation}
We suppose that $\phi_2$ condenses at some high energy scale, $\eta$,
whereas other interactions occur at a much lower scale (and will be ignored). 
This model has stable
topological strings classified by the winding number of the second
Higgs field. The reason for calling this a pruned electroweak string
should now be apparent. Although the $\phi_2$-condensation ensures that
there is no SU(2) symmetry, the solution for a
static straight string of winding number N in cylindrical polars,
\begin{eqnarray}
\Phi &=& {\eta \over \sqrt 2} \pmatrix{ 0\cr X(\rho) e^{iN\phi}\cr},\\ 
Z_\mu &=& \cos \theta_w B'_\mu - \sin \theta_w B_\mu ={2 N\over \alpha} (1 -
P(\rho) ) \partial_\mu \phi , \\
A_\mu &=& \sin \theta_w B'_\mu +\cos \theta_w B_\mu =0 , 
\end{eqnarray} 
(where $X$ and $P$ are the Nielsen-Olesen profiles, and 
$\sin \theta_w = g/\alpha $, $\alpha^2 = g^2+g^{\prime2}$) is
in fact identical to that of the electroweak Z-string (with $g, \ g'$
interchanged)\cite{N,V}. However, unlike the electroweak string, this
vortex is topologically stable.

As in the Weinberg-Salam model, define $e = gg'/\alpha$. The generators
of the broken ($\chi$)  and unbroken ($\psi$) U(1)'s in terms of the original U(1) generators
($\theta$, $\theta'$) are 
$\chi = \sin\theta_w \theta - \cos\theta_w \theta' $ and 
$\psi = \sin\theta_w \theta' + \cos\theta_w\theta $, thus
\begin{equation}
\phi_1 \to e^{i(e\psi - {\alpha\over2}\cos2\theta_w\chi)}\phi_1 \ , \ \ 
\phi_2 \to e^{i{\alpha\over2}\chi}\phi_2
\end{equation}

Now consider ${\bf g}$ given in ($\chi, \psi$) coordinates by ($
{4\pi\over\alpha}, 0 $), or in ($\theta,\theta'$) coordinates by
(${4\pi g\over\alpha^2}, - {4\pi g'\over\alpha^2}$).  Clearly, since
$H$=U(1), {\bf g} is connected to the identity, 
and indeed the rotation $\Delta\psi = {4\pi\over\alpha}
\tan\theta_w$ will return {\bf g} to the identity. 
Hence in principle
these strings can split by nucleating charged black holes. To decide
what charge the monopole must have, we can appeal to the DQC for the
symmetric phase (that is, before the phase transition). The $U(1)_g$
magnetic charge must be a multiple of $1/g$, $q=n/g$, and the
$U(1)_{g'}$ magnetic charge a multiple of $1/g'$, $q' = n'/g'$.  Since
we know that the Z-flux is $4\pi N/\alpha$, this implies that the
Z-projection of magnetic charge, $q_Z = q'\cos\theta_w -
q\sin\theta_w$, must be $N/\alpha$, i.e.\ $n'-n=N$. This gives an
A-magnetic charge of
\begin{equation}
q_A = {n+n'\over 2e} + {N(g^2-g^{\prime2})\over 2e\alpha^2}
\end{equation}
consisting of a piece satisfying the DQC (with respect to the A-electric charge $e$)
plus a non-trivial remaining part.
 
Note that if ${2e\over\alpha}\tan\theta_w = \sin2\theta_w\tan\theta_w \notin
{\bf Z}$, then $\cos2\theta_w\notin {\bf Z}$ hence ${\bf g} \phi_1\neq\phi_1$
and we have Aharonov-Bohm interactions of the $\phi_1$ quanta with the
$\phi_2$-string ({\it despite} $H$ = U(1) being connected).
The reason such a string can break is precisely that the
monopoles nucleated do not satisfy the electromagnetic
DQC and hence are naturally associated
with `visible' (non) Dirac strings. The Aharonov-Bohm phase  simply 
gets ``transferred'' from the string to the monopoles.
 
Suppose  instead we want to split the string by uncharged black
holes. This we can only do if ${\bf g} = 1$ i.e.\ if $\Delta\psi =
2\pi n / e $ for some $n\in {\bf Z}$, i.e.\ if $\tan\theta_w = {\alpha
n\over 2e}$. This requires $g = 0$, $g'=0$, or $g = g'$.  Note again
that, for all $g, \ g'$, a ${\cal G}(\phi)$ can be defined which would 
unwind the string, but is not closed in $G$.

It is obvious that these results generalize to electroweak strings: they 
can split by nucleating regular monopoles, which may then 
collapse to form magnetically charged black holes; but they can only 
split by neutral black holes when their Z
flux satisfies the DQC, that is, if $2N sin^2 \theta_W$ is an 
integer, where N is the winding number of the vortex
(which is  not a topological invariant in this case).

Now let us return to the original U(1) $\times$ U(1) $\to$ U(1)
C-metric instanton of references \cite{EHKT,HR,E1}. The magnetic
charge of the black hole in the C-metric is, in principle,
arbitrary. However, it is interesting to note that adding a particle
which is neutral under the broken U(1) (thereby having no
Aharonov-Bohm interaction with the string), and charged under the
unbroken U(1), has the effect of singling out a {\it discrete} set of
allowed values of the black hole charge -- those that satisfy the DQC
-- even though $\pi_0(H)$ is unaffected by the presence of the new
particle, or by the value of its charge. Once again we conclude that
the topological $\pi_0(H)$ criterion has to be supplemented by a
careful analysis of the DQCs.
 
Finally, we would like to address an interesting point raised by these
examples. Generally, when one talks of a tunnelling process, one is
usually performing a semiclassical approximation, using a classical
euclidean solution - the instanton - as a saddle point in the
functional integral. In each case considered here, we can still define
a euclidean solution by simply embedding the abelian Higgs instanton
in the larger theory. However, if the DQC's of the larger theory are
not satisfied, the putative instanton  is `isolated' in field
configuration space because small perturbations in certain directions
cannot be defined. In that case it is no longer a viable saddle point
and perturbation theory has broken down. Even so, it is reassuring (although
perhaps misleading) that 
the path integral calculation can still yield
the correct result -- if one interprets
the cancellation of the tunnelling amplitude as coming from the
prefactor by virtue of the fact that the only
allowable perturbations around the putative instanton form a set of
measure zero. Nevertheless, the fact that the euclidean solution still
appears to exist is deceptive, and does not indicate the presence of
an approximate solution to the full quantum theory.

%%%%%%%%%%%%%%%%%%%%%%%%%%%%%%%%%%%%%%%%%%%%%%%%%%%%%%%%%%%%%%%%%%%%%%%%
\section*{Acknowledgments}
 
Our thanks to Peter Bowcock, Ed Corrigan, Roberto
Emparan, Jeff Harvey, Mark Hindmarsh and Maxim Mostovoy for 
useful discussions and suggestions. We are grateful to the Isaac
Newton Institute for their hospitality and to the European
Commission's HCM program for financial support through contract
CHRX-CT94-0423.  A.A.\ was partially supported by grants NSF
PHY-9309364, UPV-EHU 063.310-EB225/95 and CICYT AEN-96-1668. R.G.\ was
supported by a Royal Society University Research Fellowship.

\end{document}